\documentstyle[12pt,epsfig,aaspp4]{article}
\newcommand{\be}{\begin{equation}}
\newcommand{\ee}{\end{equation}}
\newcommand{\ba}{\begin{eqnarray}}
\newcommand{\ea}{\end{eqnarray}}

\newcommand{\gsim}{\mathrel{\hbox{\rlap{\lower.55ex \hbox {$\sim$}}
                   \kern-.3em \raise.4ex \hbox{$>$}}}}
\newcommand{\lsim}{\mathrel{\hbox{\rlap{\lower.55ex \hbox {$\sim$}}
                   \kern-.3em \raise.4ex \hbox{$<$}}}}

\begin{document}

\lefthead{Lee, Brown, Wijers}

\righthead{Issues in the Blandford-Znajek Process}

\title{Issues in the Blandford-Znajek Process for GRB
Inner Engine}

\author{Hyun Kyu Lee$^{a,b}$, G.E. Brown$^{b}$,
   and R.A.M.J.  Wijers$^{b}$}
\affil{ a) Department of Physics, Hanyang University, Seoul 133-791, Korea\\
       b) Department of Physics and Astronomy,
        State University of New York at Stony Brook\\
        Stony Brook, New York 11794-3800, USA}

\begin{abstract}
Several issues regarding the Blandford-Znajek process are
discussed to demonstrate that it  can be an effective mechanism
for powering gamma-ray bursts. Using a  simple circuit
analysis it is argued that  the disk power increases the effective
power of the black hole-accretion disk system, although a part of
disk power can be dissipated into black hole entropy. Within the
framework of a force-free magnetosphere with a strong magnetic
field, a magnetically dominated MHD flow is found to support the
Blandford-Znajek process and it is demonstrated that the  possible
magnetic repulsion by the rotating black hole will not affect the
efficiency substantially.
\end{abstract}

\newpage
Recently there has  been increasing interest on the Blandford-Znajek
process (Blandford and Znajek 1977; Thorne, Price and MacDonald 1986)
as one of the viable models of powering the gamma-ray bursts
effectively (Lee,Wijers and Brown 1999).
The global picture
of magnetosphere around the rotating black hole suggested by Blandford and
Znajek is the force-free configuration of
electromagnetic charge and
current distributions (MacDonald and Thorne 1982; Okamoto 1992)
with the rotating magnetic field lines .
The magnetic field is supposed to be supported by the magnetized accretion
disk.

To set the Blandford-Znajek process (Lee,Wijers and Brown 1999)
working to  extract the rotational energy of the black hole, the
magnetic field lines and the currents should be anchored onto the
horizon (or effectively onto the stretched horizon  (Thorne, Price
and MacDonald 1986)). The toroidal magnetic field, $B_{\phi}$,  on
the horizon together with the perpendicular component, $B_H$,  to
the horizon (or on the stretched horizon) are responsible for the
outward Poynting flux in the Blandford-Znajek process which
extracts the rotational energy of the black hole. The boundary
condition on the horizon (Znajek 1977), $B_{\phi} = (\Omega_F -
\Omega_H)\tilde{\omega} B_H$\footnote{$\Omega_F$ and $\Omega_H$ are 
the angular velocities of the rotating black hole and the magnetic
field lines respectively. $\tilde{\omega}$ is the `cylindrical radius'
in the Kerr metric;~~$g_{\phi\phi} = \tilde{\omega}^2$.}, 
implies that the poloidal currents
onto the horizon are essential both to the toroidal and the
perpendicular components of the magnetic field  on the horizon.
Therefore it is an  interesting question to ask whether the
current flows onto the horizon can be realized in the relativistic
magnetohydrodynamic consideration, since
 the currents
 consist
of charged particles.

The magnetic field  on the black hole cannot
be supported by the black hole alone. The
accretion disk surrounding  the black  hole is
the  natural candidate  for the supporting
system of the strong magnetic field on the black hole.
Therefore  the disk can put some
constraints on  the
Blandford-Znajek  power.
One of the simple ways of analyzing the effect of the disk is
to adopt a   circuit analogy (Thorne, Price and MacDonald 1986; Li 1999).
The currents  out of
the
black hole in the equatorial plane  may either go directly to the
loading regions or
pass through
the accretion disk to make the closed circuit of currents
together with the inflowing currents
onto the horizon from  the loading region  at infinity.
In general,  there is also the power from  the
magnetic braking from  the accretion  disk, which extracts the disk energy and
angular momentum  out to the loading region.
Hence it increases the
power of the black hole - accretion disk system, although  a part of it
is  dissipated into
the black hole, in some cases more than the black hole power of
the magnetic braking (Li 1999).

The powers from the disk and the  black hole depend on the magnetic
field strengths.
Simple arguments using   Ampere's law and  the boundary condition  on the black
hole show  that the  poloidal magnetic field on
 the black hole is larger than the toroidal magnetic field  on the inner edge of 
the accretion disk ($r_{in}$)
\ba
B_H \gsim 2 B_{\phi}^{disk}(r_{in}).
\ea
Similar observations can be found in Ghosh and Abramowicz (1997).
Adopting the force-free field configurations suggested by Blandford
(1976) or  by  Ghosh and Abramowicz (1997), one can show that $B_H$ is
larger than
the poloidal magnetic field on the disk (Lee, Wijers and Brown 1999),
which implies   that the Blandford-Znajek power is not dominated  by
the disk power due to the magnetic braking (Li 1999, Livio, Ogilvie and
Pringle 1999). For models in which the disk field generated by disk
dynamo effects and turbulent (e.g., Balbus and Hawley 1998) it has been
argued that no more energy can be released than the disk binding energy,
and that any BZ power from the black hole is smaller than that from the
disk (Livio, Ogilvie, and Pringle 1999). Lee et~al.\ (1999) showed that
for other field configurations, such as that of Blandford (1976), or such
as may result from disruption of a neutron star around a black hole,
neither of those two limits need apply.

In this note, we discuss  the issues of the  effect of the disk
on the Blandford-Znajek
process using a  simple circuit analysis  and also of
the  magnetic field  onto the horizon of
the black  hole in the frame work of MHD
(Takahashi et al. 1990; Hirotani et al. 1992)  to show that the
Blandford-Znajek process from the system of  black hole and
the  accretion disk can
be effective enough
to power the observed GRBs.

The Poynting flux out of the horizon is carried
out to the load region along the poloidal
magnetic field lines.
Consider the  Poynting flux  through the  funnel
between  the magnetic  surfaces with
magnetic flux  $\Delta \Psi$(Thorne, Price and MacDonald 1986).
 One can   replace the
complexity of the load region   by the equivalent resistance
$\Delta R_L$ which can be formally defined by 
\ba \frac{1}{2\pi}
\Omega_F \Delta\Psi = I \Delta R_L, \label{RL} 
\ea 
where $I$  is
the   current along the   magnetic field line   and $\Delta R_L$
is the equivalent resistance   across the  load region   at
infinity  encompassed by   two magnetic surfaces  with the
magnetic flux $\Delta \Psi$. In the circuit analogy, the  left
hand side of eq.(\ref{RL}) is the voltage drop, $\Delta V_L$,
across the load region. On the horizon the magnetic braking
induces the electromotive force $\Delta {\cal E}_H$,
\be
\Delta {\cal E}_H = \frac{1}{2\pi} \Omega_H \Delta\Psi, \label{EH}
\ee
and the dissipation $\Delta P_H$ into the  horizon to increase
the entropy, $S$, of the black hole with temperature $T_H$(
$\Delta P_H \Delta t = T_H \Delta S$),
which can be written as
\ba
\Delta P_H &=& \frac{(\Delta V_H)^2}{\Delta R_H}, \label{PH}
\ea
using  the voltage drop $\Delta V_H$ on the horizon between the magnetic
surfaces ( $ \Delta
V_H = I \Delta R_H$) and the horizon resistivity $R_H$.
   The electromotive force is the sum of the
voltage drops, $\Delta V$,  across the horizon and the load region
in the circuit, which
can be written as
\ba
\Delta V = \Delta V_L + \Delta V_H
\ea
and the power delivered to the load region by the Poynting flux is
\be
\Delta P_L = I^2 R_L =
(\frac{\Delta V}{\Delta R_L+\Delta R_H})^2 \Delta R_L, \label{PL}
\ee
The Blandford-Znajek process is a sum of these circuits over
the magnetic surfaces.
For a simple analysis we  will consider the equivalent  circuit
for the Blandford-Znajek
power, where   the  electromotive forces are  summed to be
an electromotive force
${\cal E}_H$: The  current $I$  (total current flowing  into the black
 hole), the  total
horizon surface resistance $R_H$  and the equivalent load  resistance
$R_L$(Fig. 1(a)). Then from eq.(\ref{PL})
the BZ power out of the rotating black hole can be written as
\ba
P^H_{BZ} = I_H^2 R_L = (\frac{V_L + V_H}{R_L+R_H})^2{R_L}
, \label{phbz}
\ea
where the electromotive force of the black hole ${\cal E}_H = V_L + V_H$.
Since  the magnetic field lines do not  cross each other there are
no interference effects
between the magnetic fields and we may simply sum the contributions from
all field lines.

A similar analysis can be applied to the  Poynting power from
the magnetic braking of  the
magnetized accretion disk (Blandford 1976).  For the case that there
are no magnetic field lines
from the black hole  horizon (no magnetic braking of  the black hole),
the load region is
anchored by the magnetic  field lines from the  disk.
 The electromotive  force ${\cal E}_D$ on the
disk with the resistance $R_D$   and the  load region resistance, $R_L$,
  establish
a circuit (Fig. 1(b)). Then
the power at the load region is given by,
\be
P^D = I_D^2 R_L = (\frac{V_D + V_L}{R_L + R_D})^2 R_L
,\label{pd}
\ee
where $V_D$ is the voltage drop across the disk
with effective resistance $R_D$,
\ba
V_D &=& I_D R_D, \\
{\cal E}_D &=& V_D + V_L. \label{E_D}
\ea
If the circuit of current flows includes
the  horizon from  near to the  black hole's cap to the equator,
then the dissipation due to the horizon resistivity should be taken into
account.  Then
\be
{\cal E}_D =  V_L +  V_D +  V_H
\ee
and the power at the load region becomes
\be
P^{DH}
= (\frac{V_L + V_D + V_H}{R_L+ R_D + R_H})^2R_L.
 \label{pdh}
\ee

Now considering the  magnetic fields both on the horizon of
the rotating black  hole and on the accretion
disk, there are two electromotive  forces
${\cal E}_H$ and ${\cal E}_D$(Fig. 1(c)).
 The load region
where the magnetic field lines are  anchored is now divided into
two parts, of which effective resistance can be denoted by $R_L^H$ and $R_L^D$
respectively   according to
the origin of the magnetic field
lines (from the horizon or  the accretion disk).   The
magnetic field lines do not  cross each other and   we can  sum the
powers from
each region to get the  total power.  Since it  is equivalent
to the  series connection of
two resistances, we can take $R_L = R_L^H + R_L^D$ for the total power.
Then we have
\ba
{\cal E}_H+{\cal E}_{D} &=&  V_L + V_D + V_H, \\
V_L &=& I_{HD} R_L
\ea
and the power at the load region is given by
\ba
P^{HD}_{BZ} = I_{HD}^2 R_L  = (\frac{V_L + V_D + V_H}{R_L + R_D + R_H})^2R_L
\label{phdbz}
\ea
This is the power at the loading  region powered by the system of   the rotating black
hole  and the magnetized accretion disk.  Assuming high
conductivity of the accretion disk we can take $R_D$ to be negligible compared
to $R_H$ and $R_L$($V_D \sim 0$).
It is easy to see that the power is enhanced by the
addition of the accretion disk:
\be
\frac{ P^{HD}_{BZ} }{ P^{H}_{BZ}} =
 (1+ ({\cal E}_D/{\cal E}_H))^2 > 1,\label{ratio1}
\ee
Taking ${\cal E}_D \sim {\cal E}_H$,
the power from the black hole - accretion disk circuit
 would give $\sim 4$ times of that
through the black hole alone.
It is also interesting to compare $P^{HD}_{BZ}$ with the power
of the load region with
the disk battery only, $P^D$,  in order to see how  the power is
constrained by  the disk
power.  From eq.(\ref{pd}) and eq.(\ref{phdbz}) we get
\be
\frac{ P^{HD}_{BZ} }{ P^{D}}=
(\frac{R_L}{R_H +R_L})^2 (1+ ({\cal E}_H/{\cal E}_D))^2. \label{hdd}
\ee

Since the electromotive forces are induced by the magnetic braking, the ratio
${\cal E}_H/{\cal E}_D$ depends on  how to estimate   the
magnetic fields on the disk and  the horizon.   The comparison of these two
powers is possible because the magnetic fields on the black hole and the disk
are supposed to be related. The  estimation in (Li 1999) gives a
bound
\ba
{\cal E}_H/{\cal E}_D &<&  1.7,
\ea
and for the optimal case
\ba
2.5 &<& \frac{ P^{HD}_{BZ} }{ P^{H}_{BZ}}, \\
\frac{ P^{HD}_{BZ} }{ P^{D}} & < & 1.8.
\ea
On the other  hand, the estimation  using the
boundary conditions  on the  horizon and  the  disk leads  to
different results (Lee, Wijers and Brown 1999).  For
example,   for  $\tilde{a} \sim .5$ we get
\ba
 {\cal E}_H/{\cal E}_D \sim 3, \\
\frac{ P^{HD}_{BZ} }{ P^{H}_{BZ}} \sim 1.7, \\
 \frac{ P^{HD}_{BZ} }{ P^{D}} \sim 4 \label{ratio}
\ea

One can see that   the difference in the estimates is found to be 
only a factor 2 (not an
order of   magnitude or  more)  for the   moderate range  of
the angular   momentum
parameter $0.5 < \tilde{a} < 1$.
Hence the power loss from the disk into the black hole may not be a
 serious problem  in
powering GRB  at the load region as shown in eq.({\ref{ratio1}).

In this analysis we have assumed an
ideal efficiency of the disk power. In the realistic case,
the effective electromotive force ${\cal E}_D^{eff}$ should be smaller than the
total magnetic braking power.
 It is also interesting to note that a part of the disk power may come from
the rotational energy of the black hole due to the possible magnetic coupling
of the black hole and accretion disk around the equatorial plane (Gammie 1999).
Hence it is not  clear how the efficiency of the Blandford-Znajek process
 can be affected substantially by the accretion disk.

In the Blandford-Znajek process, the energy and the angular
momentum flows along the poloidal field lines are dominated by the
magnetic field. However, the magnetosphere proposed by Blandford
and Znajek should be supported by the currents, which are carried
by the fluid particles although they carry only negligible energy
and  angular momentum compared to the Poynting flux. Therefore it
is important to see whether it is dynamically possible that the
fluid particles can flow onto the horizon of the black hole, since
the currents anchored to the horizon are essential for the
Blandford-Znajek process to work in extracting the rotational
energy of the black hole. The question  on whether the current
flows in  the Blandford-Znajek   process can be realized in the
relativistic MHD consideration is related also to the  issue  of
magnetic flux repulsion due to the black hole rotation.

Assuming a steady and  axially symmetric  magnetosphere around  the
rotating black  hole, Takahashi  et~al.\ (1990) and
 Hirotani et~al.\ (1992)
discussed the physical  characteristics of the  flow along the
poloidal magnetic field  lines. The particle flows are constrained
by several critical points, which the fluid particles should
cross without causing any divergences to get onto the horizon or
 to infinity.
The poloidal velocity of the fluid particle, $u_p$,
and the derivatives along the field line
define    the      Alfven    point    and
the    fast    magnetosonic     points
respectively (Takahashi et al. 1990).
It has been shown that the negative
energy flow into the horizon is possible if
\be
0 < \Omega_F < \Omega_H, \label{ofh}
\ee
and  if the Alfven points are inside the ergosphere. This same condition
also defines the range of $\Omega_F$ for which energy is extracted 
out of the rotating black
hole via the Blandford-Znajek process.
The fluids which pass through the Alfvenic point inside the ergosphere  with
negative energy and angular momentum  should cross the fast magnetosonic point
to fall into the horizon. Hirotani et al.(1992) investigated the
case with  the fast  magnetosonic point  very near  to  the horizon
to show  that the
magnetically dominated MHD flow, which supports the poloidal currents onto
the horizon
for the Poynting flow in the Blandford-Znajek process, is possible.

There have been a series of works (Wald 1974; King, Lasota and Kundt 1975; Bicak
and Janis 1985)
which seem to imply that magnetic
flux repulsion can suppress the efficiency of the Blandford-Znajek process
substantially for the extreme Kerr black hole\footnote{More discussions with
a different   point of view
for the force-free
environment can be found in Lee, Wijers  and Brown 1999.}.
The effect of the flux expulsion
when the rotating black hole is immersed in the external uniform  magnetic
field along the rotational axis of the black hole can be expressed as
\ba
\Phi_{\tilde{a}} = ( 1- \tilde{a}^4)\Phi_{\tilde{a}=0}, \label{expel}
\ea
 which shows there is no magnetic field flux through the black hole (half
hemisphere) for the
extremely rotating  black hole, $\tilde{a}=1$.
One can see easily that the effect
gets reduced
rapidly as $\tilde{a}$  decreases.   For example if we take  
$\tilde{a}=0.5$, then
$\Phi_{0.5}/ \Phi_{0} = 0.94$. Hence for the practical
purpose of explaining the gamma-ray burst power, where the black
holes in the center are rapidly rotating but not necessarily at
the extreme rotation, it is not a severe restriction.

  In summary, it is most likely that off the equatorial plane the
Blandford-Znajek process is at least
as effective as in the original formulation,
 which has  enough efficiency for powering
gamma-ray bursts,  provided that there is a  strongly magnetized
 accretion disk.
It is also interesting to note that the interaction between the
black hole and the accretion  disk is also important, since it may affect
the  Blandford-Znajek process\footnote{For example, the effect of the 
accretion on the Blandford-Znajek power evolution has been discussed recently
by Moderski, Sikora and Lasota(1998) and Cavaliere and Malquori(1999) for 
the systems with supermassive black holes.}.
The   transition region between the horizon and the
inner edge  of the accretion disk is quite complicated compared to
the off equatorial plane up to  near the rotational  axis of the
black hole. The possible magnetic coupling between the  horizon
and the  accretion disk may  result in constraints on the
rotation of the black hole and the accretion disk. 
The effect of a magnetic field on the matter in the
accretion disk considerably complicate the discussion, but some
recent attempts have been made to incorporate this into the models
(Punsly 1998, Krolik 1999, Gammie 1999).

\acknowledgements
 We are indebted to  Li-Xin Li  for many in-depth
discussions. We also would like to thank
J. Bicak, B. Punsly, and M. Takahashi
for useful discussions.  We would also like to thank an anonymous referee 
for valuable comments. This work is supported partially
by the U.S. Department of Energy Grant No. DE-FG02-88ER40388
and also in part by
the KOSEF Grant No. 1999-2-112-003-5 and  by BSRI 98-2441.

\newpage

\begin{figure}
%\plotone{hkhk1.eps}
\begin{center}
\epsfig{file=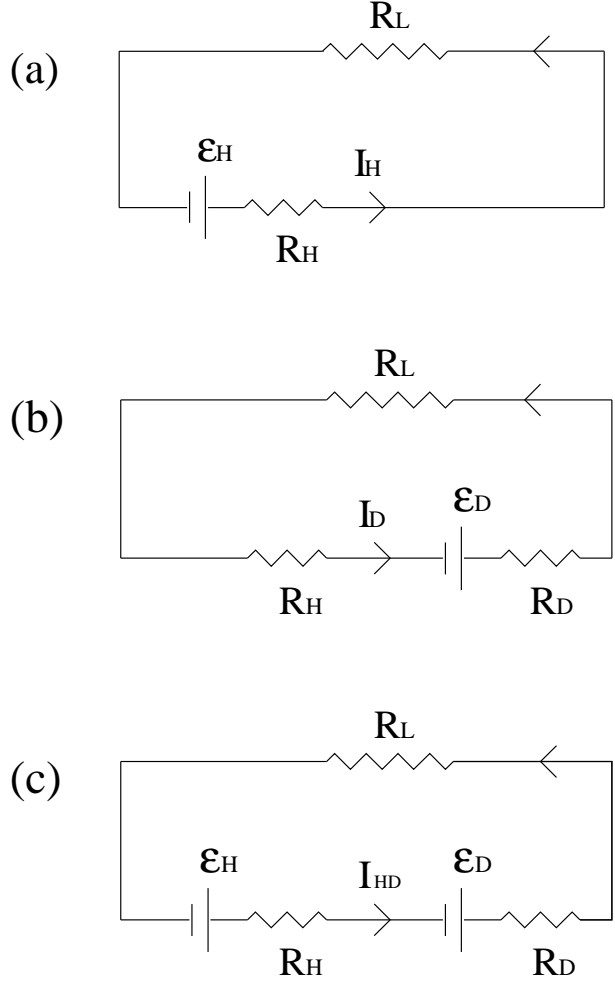, height=13cm}
\caption{
 Three  schematic circuits corresponding to
the  system of  the rotating black hole-accretion
disk in a force-free magnetosphere. The load region is
approximated by the resistance $R_L$. The electromotive forces (resistances)
 of the black
hole and the accretion disk are denoted by ${\cal E}_H(R_H)$ and
${\cal E}_D(R_D)$
 respectively. The currents $I_H, I_D$ and $I_{HD}$ are flowing out of
the black hole into the accretion disk and then to the loading region
coming back to the black hole to complete the circuit.
(a) A system of the rotating black hole alone. (b) A system of
the black hole-accretion disk
but without the Blandford-Zanjek power from the black hole. (c) A system of
the black hole - accretion disk with both the Blandford-Znajek power of the
black hole and the accretion disk.}

\end{center}
\end{figure}


\newpage

\begin{thebibliography}{}

\bibitem{} Balbus, S. A. and Hawley, J. F. 1998, Rev. Mod. Phys., 70, 1

\bibitem{} Bicak, J. and Janis, V 1985, MNRAS,  212, 899

\bibitem{} Blandford, R.D. 1976, MNRAS, 176, 465

\bibitem{} Blandford, R.D. and Znajek, R.L. 1977, MNRAS, 179, 433

\bibitem{} Cavaliere, A. and Malquori, D. 1999, ApJ, 516, L9 

\bibitem{} Gammie, C.F. 1999, ApJ. 522, L57

\bibitem{} Ghosh P. and  Abramowicz M.A. 1997, MNRAS, {\bf 292}, 887


\bibitem{} Hirotani, K., Takahashi, M.,  Nitta, S., and Tomimatsu, A. 1992,
ApJ, 386, 455


\bibitem{} King, A.R., Lasota, J.P. and Kundt, W. 1975, Phy. Rev., D12,
3037

\bibitem{} Krolik, J.H. 1999, ApJ, 515, L73


\bibitem{} Livio M., Ogilvie G.I. and  Pringle J.E. 1999, ApJ {\bf 512}, 100

\bibitem{} Lee, H.K.,  Wijers, R.A.M.J.,  and  Brown, G.E. 1999,
 Phys. Rep. to be
published, astro-ph/9906213

\bibitem{} Li, L.-X. 1999, astro-ph/9902352

\bibitem{} Macdonald, D. and Thorne, K.S. 1982, MNRAS, 198, 345

\bibitem{} Moderski, R., Sikora, M. and Lasota, J.-P. 1998, MNRAS, 301, 142 

\bibitem{} Okamoto, I. 1992, MNRAS, 294, 192

\bibitem{} Punsly, B. 1998, ApJ, 506, 790

\bibitem{} Takahashi, M.,  Nitta, S. Tatematsu, Y.,  Tomimatsu, A. 1990, ApJ,
363, 206

\bibitem{} Thorne, K.S., Price, R.H.,  and MacDonald, D.A. 1986, Black Holes;
The Membrane Paradigm (Yale University Press, New Haven and London)

\bibitem{} Wald, R.M. 1974, Phys. Rev., D10, 1680


\bibitem{} Znajek, R.L. 1977, MNRAS, 179, 457

\end{thebibliography}
\end{document}